\documentclass[]{aa}  

\usepackage{graphicx}
\usepackage{txfonts}
\usepackage{xcolor}
\newcommand{\norm}[1]{\left\lVert#1\right\rVert}

\begin{document}

  %  \title{Learning to do multiframe blind deconvolution of solar images}
   \title{Real-time multiframe blind deconvolution of solar images}

   \author{A. Asensio Ramos\inst{1,2}, J. de la Cruz Rodr\'{\i}guez\inst{3}, A. Pastor Yabar\inst{1,2,4}}

   \institute{Instituto de Astrof\'{\i}sica de Canarias, 38205, La Laguna, Tenerife, Spain; \email{aasensio@iac.es}
\and
Departamento de Astrof\'{\i}sica, Universidad de La Laguna, E-38205 La Laguna, Tenerife, Spain
\and
Institute for Solar Physics, Dept. of Astronomy, Stockholm University, Albanova University Center, 10691 Stockholm, Sweden
\and
Kiepenheuer-Institut f\"ur Sonnenphysik, 79104 Freiburg, Germany
             }

   \date{}

% \abstract{}{}{}{}{} 
% 5 {} token are mandatory
 
  \abstract{The quality of images of the Sun obtained from the ground are severely
  limited by the perturbing effect of the turbulent Earth's atmosphere. The post-facto correction
  of the images to compensate for the presence of the atmosphere require the combination 
  of high-order adaptive optics techniques,
  fast measurements to freeze the turbulent atmosphere and very time consuming blind
  deconvolution algorithms. Under mild seeing conditions, blind deconvolution
  algorithms can produce images of astonishing quality. They can be very
  competitive with those obtained from space, with the huge advantage of the flexibility
  of the instrumentation thanks to the direct access to the telescope. In this
  contribution we leverage deep learning techniques to significantly accelerate the
  blind deconvolution process and produce corrected images at a peak rate of $\sim$100 images
  per second. We present two different architectures that produce excellent image
  corrections with noise suppression while maintaining the photometric properties 
  of the images. As a consequence, polarimetric signals can be obtained with
  standard polarimetric modulation without any significant artifact.  
  With the expected improvements in computer hardware and algorithms, we anticipate that
  on-site real-time correction of solar images will be possible in the near future.}
  % {bla}
  % {The stability equations of state are
  % calculated for solar composition and are displayed in the domain
  % $-14 \leq \lg \rho / \mathrm{[g\, cm^{-3}]} \leq 0 $,
  % $ 8.8 \leq \lg e / \mathrm{[erg\, g^{-1}]} \leq 17.7$. These displays
  % may be
  % used to determine the one-zone stability of layers in stellar
  % or planetary structure models by directly reading off the value of
  % the stability equations for the thermodynamic state of these layers,
  % specified
  % by state quantities as density $\rho$, temperature $T$ or
  % specific internal energy $e$.
  % Regions of instability in the $(\rho,e)$-plane are described
  % and related to the underlying microphysical processes.}
  % {Vibrational instability is found to be a common phenomenon
  % at temperatures lower than the second He ionisation
  % zone. The $\kappa$-mechanism is widespread under `cool'
  % conditions.}

   \authorrunning{Asensio Ramos et al.}
   \titlerunning{Real-time solar multiframe blind deconvolution}
   \maketitle
%
%________________________________________________________________

\section{Introduction}
Arguably the largest difficulty to face when observing the Sun from 
Earth is the perturbing effect of the atmosphere. The turbulent
variations of the index of refraction at different layers of the atmosphere
produce distortions in the images that severely reduce the quality of the
observations. Perhaps the most obvious way of avoiding this effect is to
move the telescope to space. Examples of this are Hinode \citep{suematsu08}
and/or Sunrise \citep{sunrise10}, which allowed us to have images of the solar
surface with high quality for the whole duration of the missions.
Still, ground-based telescopes hold many advantages. Instruments can be
modified and tuned online, which can help reach unprecedented levels
of detail in the investigated solar signals. Additionally, ground-based
telescopes can be made with significantly larger apertures. Large telescopes 
are complicated and heavy machines that are better operated at ground level. 

Many efforts have been put on compensating for the perturbing effect of the
atmosphere. A very successful frontline is the development of active and
adaptive optics that measure the wavefront perturbations at high time cadence
and correct it using deformable optical elements. Working at very high frequencies
(up to a few kHz), current adaptive optics (AO) systems can very well correct
for the turbulent layers closer to the telescope, remarkably enhancing
the quality of the science data. Even with such corrections, turbulence at
higher layers, that typically produce a spatially variant image motion, remains uncorrected. To
this end, multi-conjugate AO (MCAO) systems based on several deformable mirrors that 
are conjugate with the turbulence layer at different heights have been proposed.
The first tests for solar observations \citep{schmidt_mcao17} have demonstrated that this approach is able to correct a 
much larger field-of-view (FOV).

Another frontline is the development of a-posteriori image correction algorithms.
These methods are also routinely applied even in observations carried out
with AO systems. The reason is that the corrections carried out by the deformable
mirrors are often incomplete and there is still a non-negligible atmospheric residual in the observations.
Under the assumption of the linear theory of image formation, the 
perturbing effect of the atmosphere can be compensated for using optimization
methods. In such an approach, the observed image $I$ is computed from the real object, $O$ as:
\begin{equation}
  I = P(O),
\end{equation}
where $P$ is a linear operator that characterizes the instantaneous point spread function (PSF) 
of the atmosphere at every spatial position of the image. The previous equation can often be
simplified inside small FOVs (the so-called anisoplanatic patch, which share the
same PSF) to the following convolution:
\begin{equation}
  I = P * O.
\end{equation}

Since both the real image and the PSF are 
unknown, almost all methods that are currently in use or have
potential to be developed belong to the 
class of blind (or semi-blind) inversion schemes, in which
both the solar image, $O$, and the PSF, $P$, need to be 
simultaneously obtained. Only in few cases we find examples of
non-blind inversion in which the wavefront is measured and used
to infer the PSF. Blind inversion problems are always very ill-defined and
the solution strongly depends on the assumption of priors. It is therefore necessary
to add extra information to better condition the problem, and some avenues
have been tried in Solar Physics. Arguable the first
method used in the field was the speckle technique \citep{labeyrie70,vonderluhe93}, in which many 
short-exposure images (where 
the atmospheric seeing can be considered to be frozen) are obtained. This method can be 
understood as a semi-blind inversion, in which partial statistical information about the
wavefront is estimated from models (we note that it remains to be checked whether the 
speckle method can be posed as an optimization of a merit function, like the
rest of inversion scheme discussed in this section). They are later combined to estimate
the amplitude and phase of the Fourier transform of the original object. A second
method broadly used in Solar Physics is that of phase diversity \citep{paxman92,1994A&AS..107..243L,lofdahl98}.
Under the assumption of
uncorrelated Gaussian noise, \cite{paxman92} demonstrated that
%the use of two images is specially relevant because
%it allows to write down 
a proper likelihood function can be obtained in which the image $O$
does not explicitly appear and one only needs to optimize for the wavefront.
One of the most widespread application of this approach is that in which
two images (typically one in focus and the other one defocused) are used to
jointly estimate the wavefront and the original object. To
this end, the wavefront (usually defined on a circular or annular pupil) is developed 
in a suitable orthogonal
basis (usually Zernike polynomials or Karhunen-Loeve modes) and the coefficients are
obtained by maximizing the likelihood. Once the PSF is estimated, the final 
image can be obtained with a standard non-blind deconvolution. This method has been
recently used with success for correcting the data obtained with IMaX \citep{imax11} onboard 
Sunrise \citep{sunrise10}. 

\begin{figure*}
  \includegraphics[width=\textwidth]{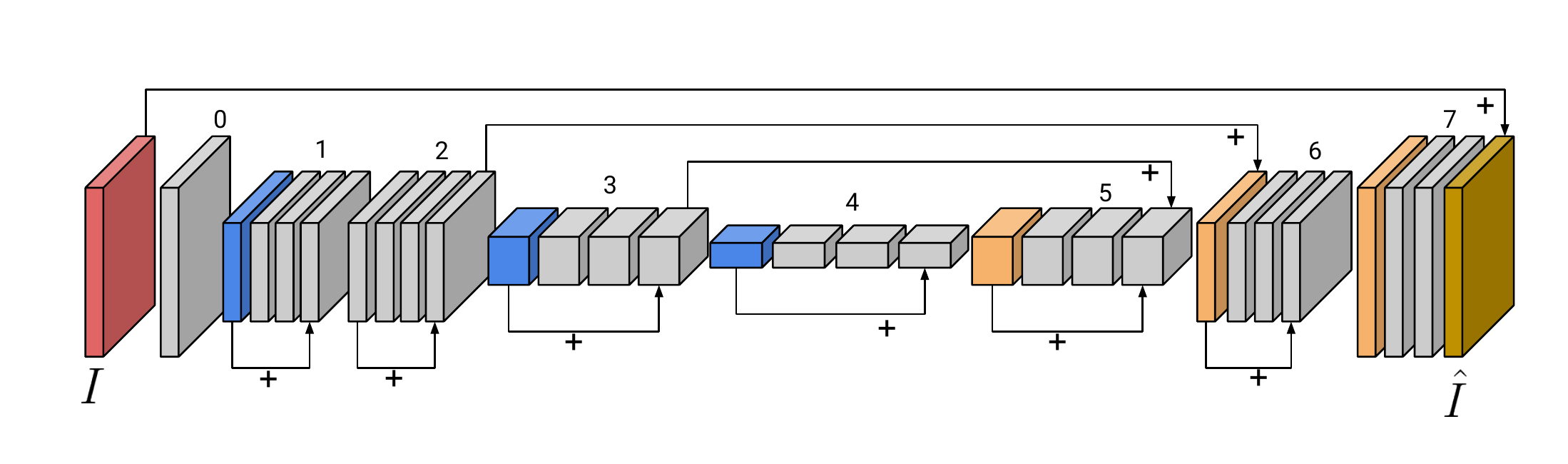}
  \caption{Architecture of the encoder-decoder deconvolution neural network. The network is composed
  of the input and 8 super-blocks, each one made of a different number of blocks. The meaning of colors
  for the blocks are described in Sect. \ref{sec:encdec}. The specific details for each block is 
  described in Tab. \ref{tab:encdec}. The numbers above the blocks label the super-blocks.}
  \label{fig:encdec_architecture}
  \end{figure*}

  \begin{figure*}
    \centering
    \includegraphics[width=0.7\textwidth]{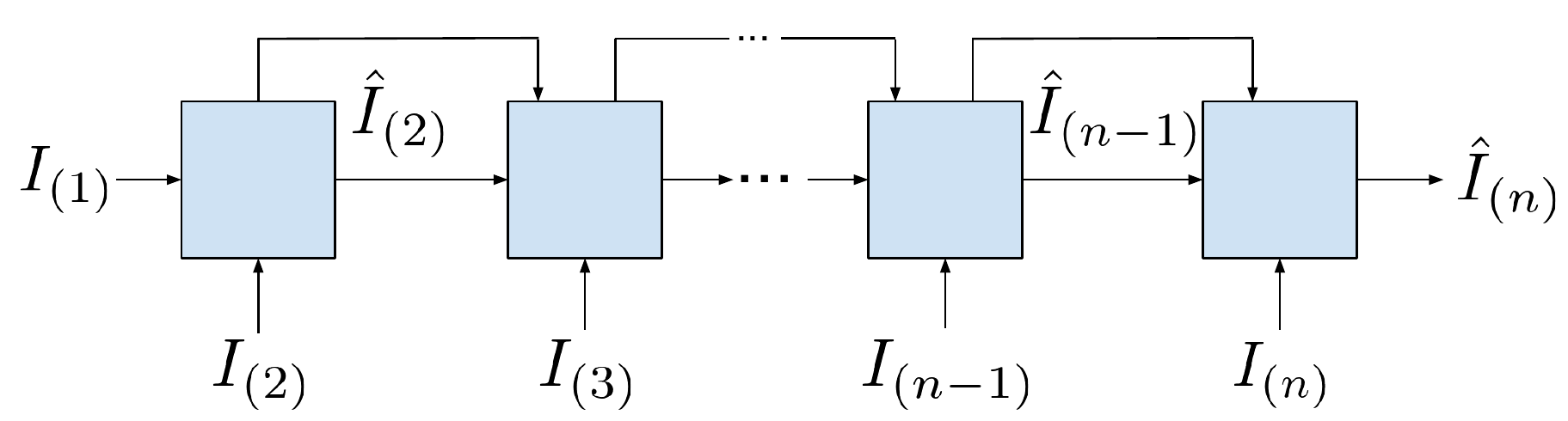}
    \includegraphics[width=\textwidth]{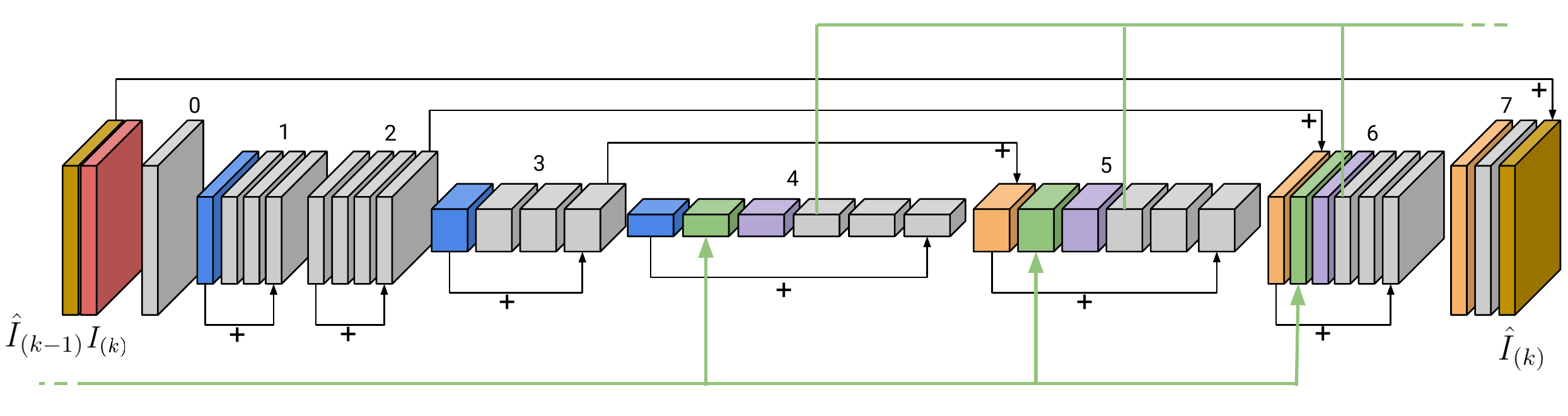}
    \caption{Upper panel: end-to-end deconvolution process, where the grey blocks are the deconvolution blocks
    described in the lower panel. Lower panel: internal architecture of each deconvolution block.
    Colors for the blocks are described in Sect. \ref{sec:recurrent}, with the specific details 
    for each block being described in Tab. \ref{tab:recurrent}.}
    \label{fig:recurrence_architecture}
  \end{figure*}

Perhaps the most successful method so far, and the one achieving the
best corrections, is the multi-object multi-frame blind deconvolution \citep[MOMFBD;][]{vannoort05}.
This method consists of a maximum-likelihood solution to the deconvolution problem
which leverage, arguably in order of importance: i) phase diversity, ii)
many frames of the same object observed with different wavefront perturbations, and iii)
many different objects (monochromatic images at different wavelengths) 
observed simultaneously with exactly the same wavefront perturbation. A method to use 
all of them was already derived by \cite{lofdahl02}, who also implemented
a computer program taking into account the first two. It was later extended to many objects
by \cite{vannoort05}. The enormous
redundancy of such the multi-object and multi-frame scheme introduces a strong
constraints in the maximum-likelihood solution and obtain excellent deconvolved images.

The fundamental problem of the majority of a-posteriori image processing algorithms
is of computational character, both 
in terms of computing power and memory. Of special relevance is the case of MOMFBD, in which large
supercomputers working during many hours are needed to deconvolve the observed data. This is a 
very limiting factor which will become even more severe with the advent of 4-m class telescopes
in Solar Physics. In this contribution we leverage end-to-end deep learning solutions
for deconvolving multi-frame bursts of solar images. The resulting
schemes can be deployed on Graphical Processing Units (GPUs) and can produce deconvolved
images almost on real-time. This opens up the possibility of on the fly blind deconvolution
in current and future telescopes. The methods that we present here can be 
applied to produce science-ready data or can be used to help the observer to 
have a better intuition of what can be expected prior to the standard 
data reduction and MOMFBD deconvolution.

\section{Neural network architectures}
The success of machine learning approaches based on deep neural networks (DNN) is
beyond doubt. Apart from the enormous amount of 
applications\footnote{See, e.g., the curation on \texttt{https://bit.ly/2ll0dQI}.} in
computer vision, natural language processing and other fields, we also
find applications in astrophysics such as the 
classification of galactic morphologies \citep{huertas-company15} or the
development of generative models to help constrain the deconvolution of images of 
galaxies \citep{schawinski17}. In the field of solar physics, 
this approach has allowed us to infer horizontal velocity fields from
consecutive continuum images \citep{2017A&A...604A..11A} and also to
simultaneously deconvolve and superresolve images \citep{2017arXiv170602933D} 
from the synoptic telescope Helioseismic and Magnetic Imager \citep[HMI;][]{Scherrer2012} 
onboard the Solar Dynamics Observatory \citep[SDO;][]{sdo2012}.

\begin{table*}
  \caption{Architecture of encoder-decoder network}
  \label{tab:encdec}
  \centering
  \begin{tabular}{c c c c c c}
  \hline\hline
  Layer & Type & Kernel size\tablefootmark{a} & Stride & Input shape\tablefootmark{b} & Output shape\tablefootmark{b} \\
  \hline
  $\color{gray} C_{0,1}$ & Convolution & $7 \times 7 \times 64$ & 1 & $N \times N \times 7$ & $N \times N \times 64$ \\
  \hline
  $\color{blue} C_{1,1}$ & Convolution & $3 \times 3 \times 64$ & 2 & $N \times N \times 64$ & $N/2 \times N/2 \times 64$ \\
  $\color{gray} C_{1,2}$-\color{gray} $C_{1,4}$ & Convolution & $3 \times 3 \times 64$ & 1 & $N/2 \times N/2 \times 64$ & $N/2 \times N/2 \times 64$ \\
  \hline
  $\color{gray} C_{2,1}$-\color{gray} $C_{2,4}$ & Convolution & $3 \times 3 \times 64$ & 1 & $N/2 \times N/2 \times 64$ & $N/2 \times N/2 \times 64$ \\
  \hline
  $\color{blue} C_{3,1}$ & Convolution & $3 \times 3 \times 128$ & 2 & $N/2 \times N/2 \times 64$ & $N/4 \times N/4 \times 128$ \\
  $\color{gray} C_{3,2}$-\color{gray} $C_{3,4}$ & Convolution & $3 \times 3 \times 128$ & 1 & $N/4 \times N/4 \times 128$ & $N/4 \times N/4 \times 128$ \\
  \hline
  $\color{blue} C_{4,1}$ & Convolution & $3 \times 3 \times 256$ & 2 & $N/4 \times N/4 \times 128$ & $N/8 \times N/8 \times 256$ \\  
  $\color{gray} C_{4,2}$-\color{gray} $C_{4,4}$ & Convolution & $3 \times 3 \times 256$ & 1 & $N/8 \times N/8 \times 256$ & $N/8 \times N/8 \times 256$ \\
  \hline
  $\color{orange} C_{5,1}$ & Up-convolution & $3 \times 3 \times 128$ & 1 & $N/8 \times N/8 \times 256$ & $N/4 \times N/4 \times 128$ \\
  $\color{gray} C_{5,2}$-\color{gray} $C_{5,4}$ & Convolution & $3 \times 3 \times 128$ & 1 & $N/4 \times N/4 \times 128$ & $N/4 \times N/4 \times 128$ \\
  \hline
  $\color{orange} C_{6,1}$ & Up-convolution & $3 \times 3 \times 64$ & 1 & $N/4 \times N/4 \times 128$ & $N/2 \times N/2 \times 64$ \\
  $\color{gray} C_{6,2}$-\color{gray} $C_{6,4}$ & Convolution & $3 \times 3 \times 64$ & 1 & $N/2 \times N/2 \times 64$ & $N/2 \times N/2 \times 64$ \\
  \hline
  $\color{orange} C_{7,1}$ & Up-convolution & $3 \times 3 \times 64$ & 1 & $N/2 \times N/2 \times 64$ & $N \times N \times 64$ \\
  $\color{gray} C_{7,2}$ & Convolution & $3 \times 3 \times 64$ & 1 & $N \times N \times 64$ & $N \times N \times 64$ \\
  $\color{gray} C_{7,3}$ & Convolution & $3 \times 3 \times 16$ & 1 & $N \times N \times 64$ & $N \times N \times 16$ \\
  $\color{olive} C_{7,4}$ & Convolution & $1 \times 1 \times 1$ & 1 & $N \times N \times 16$ & $N \times N \times 1$ \\
  \end{tabular}
  \tablefoot{
  \tablefoottext{a}{Kernel spatial size and depth.}
  \tablefoottext{b}{Image spatial size and number of channels.}
  }
  \end{table*}

\begin{table*}
  \caption{Architecture of recurrent network}
  \label{tab:recurrent}
  \centering
  \begin{tabular}{c c c c c c}
  \hline\hline
  Layer & Type & Kernel size\tablefootmark{a} & Stride & Input shape\tablefootmark{b} & Output shape\tablefootmark{b} \\
  \hline
  $\color{gray} C_{0,1}$ & Convolution & $7 \times 7 \times 64$ & 1 & $N \times N \times 2$ & $N \times N \times 64$ \\
  \hline
  $\color{blue} C_{1,1}$ & Convolution & $3 \times 3 \times 64$ & 2 & $N \times N \times 64$ & $N/2 \times N/2 \times 64$ \\
  $\color{gray} C_{1,2}$-\color{gray} $C_{1,4}$ & Convolution & $3 \times 3 \times 64$ & 1 & $N/2 \times N/2 \times 64$ & $N/2 \times N/2 \times 64$ \\
  \hline
  $\color{gray} C_{2,1}$-\color{gray} $C_{2,4}$ & Convolution & $3 \times 3 \times 64$ & 1 & $N/2 \times N/2 \times 64$ & $N/2 \times N/2 \times 64$ \\
  \hline
  $\color{blue} C_{3,1}$ & Convolution & $3 \times 3 \times 128$ & 2 & $N/2 \times N/2 \times 64$ & $N/4 \times N/4 \times 128$ \\
  $\color{gray} C_{3,2}$-\color{gray} $C_{3,4}$ & Convolution & $3 \times 3 \times 128$ & 1 & $N/4 \times N/4 \times 128$ & $N/4 \times N/4 \times 128$ \\
  \hline
  $\color{blue} C_{4,1}$ & Convolution & $3 \times 3 \times 256$ & 2 & $N/4 \times N/4 \times 128$ & $N/8 \times N/8 \times 256$ \\
  $\color{green} C_{4,2}$ & Concatenate & - & 1 & $2 \times N/8 \times N/8 \times 256$ & $N/8 \times N/8 \times 512$ \\
  $\color{violet} C_{4,3}$ & Convolution & $1 \times 1 \times 256$ & 1 & $N/8 \times N/8 \times 512$ & $N/8 \times N/8 \times 256$ \\
  $\color{gray} C_{4,4}$-\color{gray} $C_{4,6}$ & Convolution & $3 \times 3 \times 256$ & 1 & $N/8 \times N/8 \times 256$ & $N/8 \times N/8 \times 256$ \\
  \hline
  $\color{orange} C_{5,1}$ & Up-convolution & $3 \times 3 \times 128$ & 1 & $N/8 \times N/8 \times 256$ & $N/4 \times N/4 \times 128$ \\
  $\color{green} C_{5,2}$ & Concatenate & - & 1 & $2 \times N/4 \times N/4 \times 128$ & $N/4 \times N/4 \times 256$ \\
  $\color{violet} C_{5,3}$ & Convolution & $1 \times 1 \times 128$ & 1 & $N/4 \times N/4 \times 256$ & $N/4 \times N/4 \times 128$ \\
  $\color{gray} C_{5,4}$-\color{gray} $C_{5,6}$ & Convolution & $3 \times 3 \times 128$ & 1 & $N/4 \times N/4 \times 128$ & $N/4 \times N/4 \times 128$ \\
  \hline
  $\color{orange} C_{6,1}$ & Up-convolution & $3 \times 3 \times 64$ & 1 & $N/4 \times N/4 \times 128$ & $N/2 \times N/2 \times 64$ \\
  $\color{green} C_{6,2}$ & Concatenate & - & 1 & $2 \times N/2 \times N/2 \times 64$ & $N/2 \times N/2 \times 128$ \\
  $\color{violet} C_{6,3}$ & Convolution & $1 \times 1 \times 64$ & 1 & $N/2 \times N/2 \times 128$ & $N/2 \times N/2 \times 64$ \\
  $\color{gray} C_{6,4}$-\color{gray} $C_{6,6}$ & Convolution & $3 \times 3 \times 64$ & 1 & $N/2 \times N/2 \times 64$ & $N/2 \times N/2 \times 64$ \\
  \hline
  $\color{orange} C_{7,1}$ & Up-convolution & $3 \times 3 \times 64$ & 1 & $N/2 \times N/2 \times 64$ & $N \times N \times 64$ \\
  $\color{gray} C_{7,2}$ & Convolution & $3 \times 3 \times 8$ & 1 & $N \times N \times 64$ & $N \times N \times 8$ \\
  $\color{olive} C_{7,3}$ & Convolution & $1 \times 1 \times 1$ & 1 & $N \times N \times 8$ & $N \times N \times 1$ \\
  \end{tabular}
  \tablefoot{
  \tablefoottext{a}{Kernel spatial size and depth.}
  \tablefoottext{b}{Image spatial size and number of channels.}
  }
  \end{table*}

In this paper we leverage multiframe (video) correction methods which, in the deep learning 
literature, are
treated using fundamentally two approaches. The first one is to fix the number
of input frames and use them as channels in a standard convolution neural network (CNN).
The output of the CNN is the corrected frame, taking into account all the
spatial information encoded in the degraded frames. 
To process a larger number of frames, one applies the DNN in batches
until all frames are exhausted. However, fixing the number of frames in the burst can be 
seen as a limiting factor and one
would like an approach that works the same irrespectively of the number of 
frames in the bursts. This is of special relevance in the special case that adding more
frames can have a large impact on the final quality of the corrected image.

The second approach is to use a DNN with some type of recurrency, so that frames
are processed in order. New frames are injected on the network and a
corrected version of the image is obtained at the output. Introducing 
new frames on the input will slowly improve the quality of the output.
This procedure can be iterated until a good enough final image is obtained.

In this paper we have explored the two options. The case in which we fix the
number of frames of the input is an end-to-end approach based on an
encoder-decoder network that we describe in Section \ref{sec:encdec}. The case of
a recurrent neural networks is based on the very flexible strategy followed by 
\cite{WieHirSchLen17} and is explained in Section \ref{sec:recurrent}. Although
the quality of the output is similar, there are pros and cons on each
one of the architecture, which we point out in the following. For many of the
technical details, we refer the reader to our previous works \citep{2017A&A...604A..11A,2017arXiv170602933D}.

\subsection{Encoder-decoder architecture}
\label{sec:encdec}
The architecture that we use in the encoder-decoder network is displayed
in Fig. \ref{fig:encdec_architecture}. The input $I$ contains a fixed number of input frames 
(commonly known as channels) of size $N \times N$. We use 7 in our case. The input is marked 
as a light red block.
Then, several standard differentiable
operations (illustrated as colored blocks) are applied in sequence. These blocks are organized
in 8 super-blocks with different image sizes (labeled with the numbers above the super-blocks).

In summary, the network follows a standard
encoder-decoder architecture. In the encoder phase the spatial
size of the images is reduced while increasing the number of channels. In the decoder
phase, the original size is recovered by upsampling. Each colored block has the following meaning:
\begin{itemize}
\item \emph{Red blocks} are input blocks, containing all 7 frames of the burst.
\item \emph{Yellow blocks} are output blocks, containing just a single deconvolved frame (channel), that
we label $\hat{I}$.
\item \emph{Grey blocks} are standard convolutional blocks made of (in this order): a batch normalization layer 
\citep[BN;][]{ioffe_batchnormalization15}, a rectified linear unit \cite[ReLU;][]{relu10} as
activation function and a convolutional layer with kernel size and depth defined in
Tab. \ref{tab:encdec}. To keep the size of the images unchanged, the images are padded
in the borders using reflection padding.
\item \emph{Dark-blue blocks} are convolution blocks like the previous ones, but using a stride
of 2 during convolution (in other words, the convolution is carried out by
sliding the kernel on the input in steps of 2 pixels). This reduces the size of the output by a factor 2.
\item \emph{Orange blocks} are upsampling convolutional blocks made of (in order) BN, ReLU, upsampling by a factor
2 using a nearest-neighbor interpolation and finally a convolutional layer. We prefer
upsampling+convolution as a substitute of the standard transpose convolution for upsampling 
given that the latter can easily produce checkerboard 
artifacts\footnote{\texttt{https://distill.pub/2016/deconv-checkerboard/}}.
\end{itemize}
For the specific details of our implementation, we refer to Tab. \ref{tab:encdec},
where each layer is indicated as $C_{s,i}$, with $s$ referring to the super-block and $i$ to
the specific block inside each super-block.

Our architecture also contains many shortcut connections displayed as arrows 
connecting two non-consecutive blocks in Fig. \ref{fig:encdec_architecture}. 
These connection \citep{residual_network16} simply add the input to the
output, producing an important acceleration on the training by avoiding 
the effect of vanishing gradients\footnote{Gradients used during training become
exponentially small when the neural network is sufficiently deep 
(e.g., \texttt{https://en.wikipedia.org/wiki/Vanishing\_gradient\_problem}).}. The 
encoder-decoder architecture has at least two advantages as compared with the fully convolutional
architectures that kept the size of the images throughout the network
and that we used in our previous works \citep{2017A&A...604A..11A,2017arXiv170602933D}. The first one
is that the computing time is reduced because convolutions are applied over
increasingly smaller images during the encoder phase. The second one is that small kernels
(like the 3$\times$3 kernels that we use in this work) can produce much larger receptive fields
(they affect much larger regions in the input image after multiple convolutions) thanks to the 
reduction in size of the images through the encoder phase\footnote{As an example, applying one $3 \times 3$
kernel on an image produces a receptive field of size $3 \times 3$. Applying it twice 
increases the receptive field to a patch of size $5 \times 5$. On the contrary, if the image
is halfed in size between both convolutions, the receptive field increases to $7 \times 7$.}
On the contrary, the training is often more difficult because the network needs to
recognize how to generate high spatial frequency in the decoder phase from the 
combination of information in different channels.

\subsection{Recurrent architecture}
\label{sec:recurrent}
The recurrent architecture we have used is displayed in
Fig. \ref{fig:recurrence_architecture}, and is essentially the one used by
\cite{WieHirSchLen17} but with a few minor modifications. It consists of a single 
encoder-decoder network very similar to our encoder-decoder architecture. This network takes as input
two frames, $I_{(i)}$ and $I_{(i+1)}$, and produces as output a deconvolved frame
$\hat{I}_{(i+1)}$ (we remind the reader that we use the hat to denote corrected frames). The output frame
is used, together with a new frame of the burst, to produce a new estimation
of the corrected frame. This is iterated until the frames of the burst
are exhausted. The architecture also propagates some information internally from
one block to the next. 
The internal structure of each deconvolution block of the upper panel of
Fig. \ref{fig:recurrence_architecture} is displayed in the lower panel of the
same figure. 
% In total, we find 7 
% superblocks, each one made of 4 or 6 fundamental blocks. 
Some of these blocks
have been already described above, while the new ones have the following properties:
\begin{itemize}
  \item \emph{Purple blocks} are similar to standard convolutional blocks but using kernels
  of spatial size $1 \times 1$. Therefore, they compute a weighted average along the
  channel dimension.
  \item \emph{Green blocks} carry out the concatenation of two inputs with the same spatial
  dimensions and the same number of channels, producing an output with twice the number of
  channels. The memory from previous frames is passed along the green arrows to subsequent iterations 
  and merged in these green blocks. Specifically, we extract features from blocks 4, 5, and 6, and
  blend them again in the same blocks for the next iteration. These connections are
  obviously not operative for the first iteration of the deconvolution block. 
\end{itemize}

The reader can note that our architecture differs from that used by \cite{WieHirSchLen17} 
in two minor points. The first one is the size of some of the kernels. The second one is that we have used 
upsampling+convolution as a substitute of the standard transpose convolution.
We also note that, apart from the possibility of injecting an arbitrary number $n$ of frames in the
recurrent architecture, this network also allows to output $n-1$ corrected frames, which
can later be used to improve the signal-to-noise ratio.

\begin{figure*}
\includegraphics[width=\textwidth]{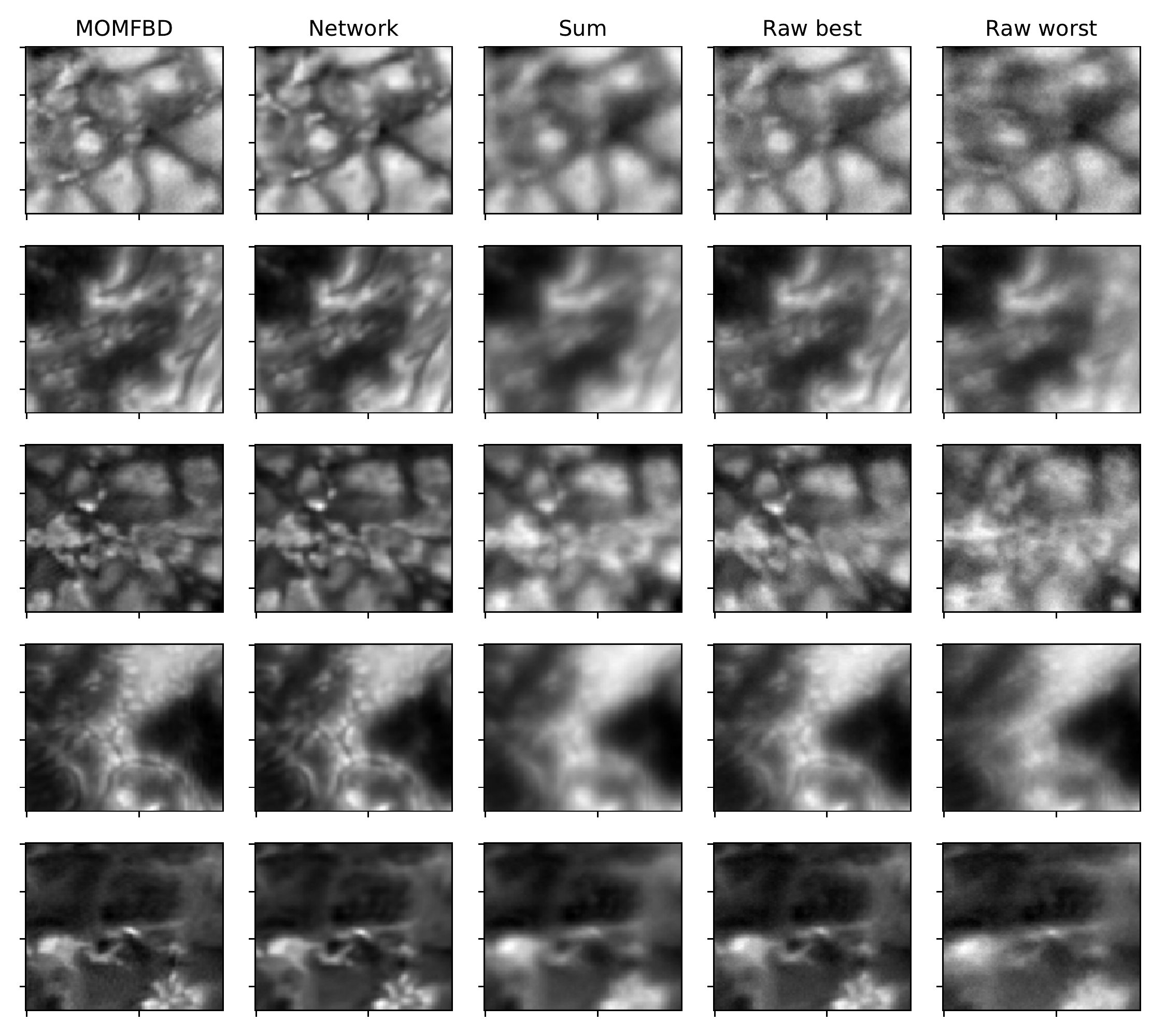}
\caption{Selection of patches from the training set. The first column shows the target
MOMFBD corrected patch. The second column shows the output ot the encoder-decoder network
when converged. The third column shows the average of the 7 frames of the burst, while
the last two columns are two of these individual frames.}
\label{fig:training_set}
\end{figure*}

\section{Training}

\subsection{Training dataset}
\label{sec:training_set}
We have trained both DNNs\footnote{The code for the networks using \texttt{PyTorch} can be downloaded from 
\texttt{https://github.com/aasensio/learned\_mfbd}.}
using two datasets observed with the CRisp Imaging SpectroPolarimeter (CRISP) instrument at 
the Swedish 1-m Solar Telescope (SST) on the Observatorio del Roque de los Muchachos (Spain). 
We have tried to use datasets that cover a broad variety of solar regions, from quiet
Sun to more active regions. The training set was obtained in fairly good 
average seeing conditions. However, there were some significant seeing variations, 
especially at the beginning of the series. Therefore, we think that the training set contains a 
broad representation of good to average seeing conditions..
One of the datasets corresponds to a quiet Sun region, observed
on 2016-09-19 from 10:03 to 10:04 UT. The other one is a region of flux emergence, observed
on the same day from 09:30 to 10:00. The data used for training are spectral
scans on the \ion{Fe}{i} doublet 
on 6301-6302 \AA, containing 15 wavelength points, and the \ion{Ca}{ii} line at 8542 \AA, 
containing 21 wavelength points. The observations are recorded in 4 modulated polarization 
states, which allow reconstructing the full-Stokes vector. CRISP includes dual beam 
polarimetry to minimize seeing-induced cross-talk \citep[see, e.g, ][]{2012ApJ...757...45C}. 
The image acquisition is performed in such a way that the four polarization states are 
interleaved sequentially until 7 acquisitions are acquired in each state. The exposure 
time per acquisition is $\sim17.35$ ms and the pixel size is $0.059"$. Additionally, strictly simultaneous wide band (WB)
images are acquired with the narrow band (NB) images, which are used for the deconvolution process.

The images are reduced following the standard procedure \citep{jaime15}, 
that includes: dark current subtraction,
flat-field correction, and subpixel image alignment between the two NB cameras 
and the WB camera. The bursts of seven images and the
simultaneous WB images are used by the MOMFBD technique to recover a deconvolved
final image. The MOMFBD code applies a Fourier filter to the reconstructed images that  
suppresses frequencies above the diffraction limit of the 
telescope (or from a practical point of view, above the noise
limit), as described in \cite{vannoort05}. We use these images as the
output of our training set. Two additional datasets of the same region obtained the same
day is used as a validation set to check for over-fitting during the training.

A total of $\mathcal{N}=80000$ patches of 88$\times$88 pixels are randomly extracted from the 
bursts of 7 images and from the
final deconvolved image. They are also randomly extracted from the spectral positions and
from the polarimetric modulation states. The size of 88 pixels allows to reduce the size of the image
three times in the encoder phase always obtaining images of integer size. Given that
the network is fully convolutional, it can be safely applied to images of 
arbitrary size. However, we point out that in order to recover an output of the
same size as the input, the number of pixels in both directions have to be
an integer multiple of 8. We also apply an augmenting strategy that consists 
of randomly flipping the patches horizontally and vertically. This improves the generalization 
capabilities of the neural networks.

\begin{figure*}
  \includegraphics[width=\textwidth]{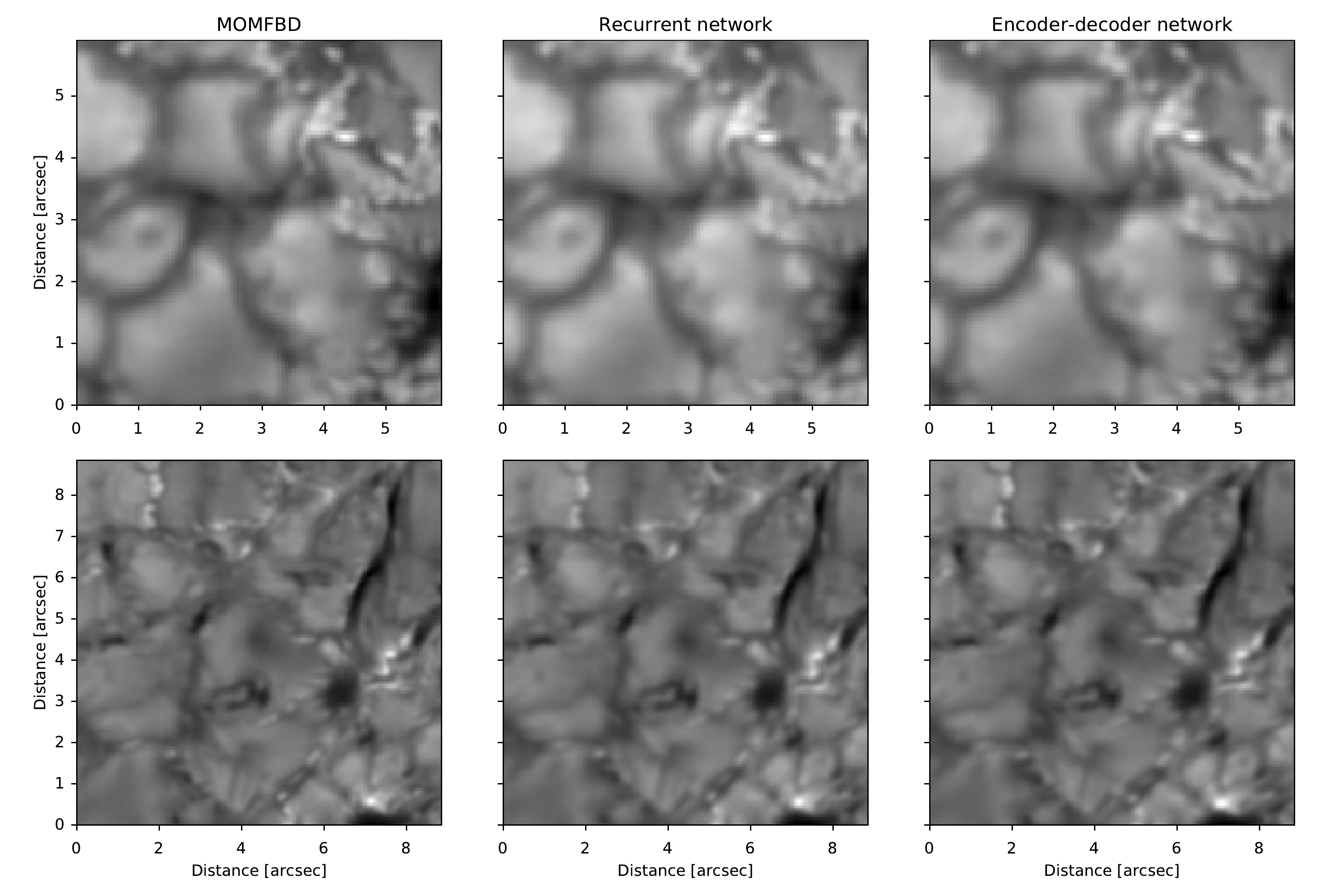}    
\caption{The first column displays the images reconstructed with MOMFBD. The central column
displays the output of the recurrent network and the right column shows the results for
the encoder-decoder network. The upper row shows an image in the continuum around
6302 \AA, while the lower panel displays an image in the core of the line in a different
part of the FOV.}
\label{fig:6302_validation}
\end{figure*}

\begin{figure*}  
  \includegraphics[width=\textwidth]{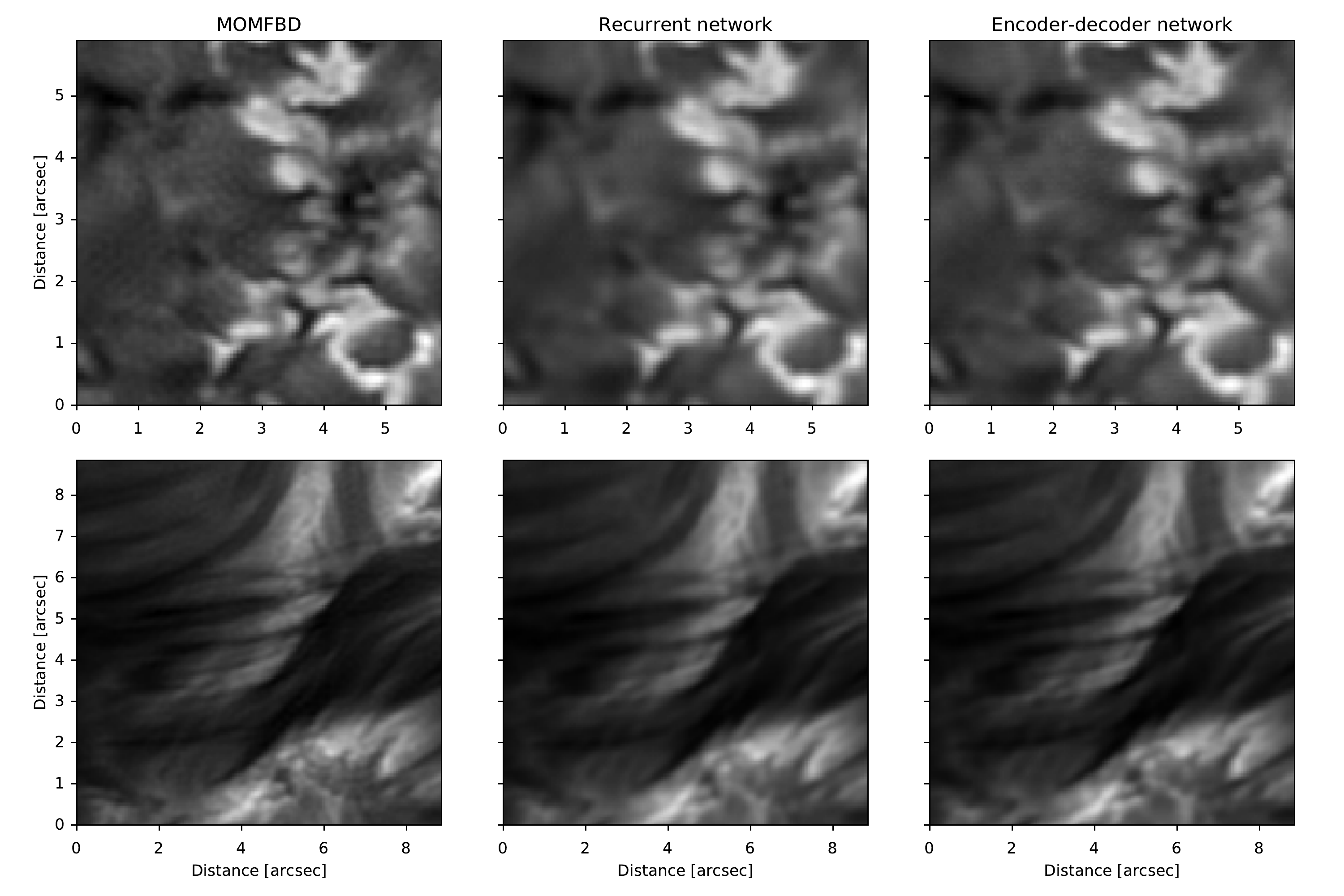}
\caption{Same as Fig. \ref{fig:6302_validation} but for the 8542 \AA\ line.}
\label{fig:8542_validation}
\end{figure*}

\subsection{Optimization}
The previous neural networks are trained by optimizing a loss
function that measures how far the outputs of the network
are when compared with the target images.
The encoder-decoder network is trained by minimizing the following loss function:
\begin{equation}
L = \sum_{i=1}^\mathcal{N} \norm{ \hat{I}(i) - I_{D}(i) }^2,
\end{equation}
that measures the sum over the full training dataset of $\mathcal{N}$ patches
of the $\ell_2$ distance between the deconvolved frames 
obtained at the output of the network, $\hat{I}(i)$, and the one deconvolved with the 
MOMFBD algorithm, $I_{D}(i)$. We note that $i$ loops indisctintly over the
randomly extracted 88$\times$88 patches, the CRISP spectral scanning positions and
the four modulation states. Summarizing, we use the scalar $L$ as a measurement of 
the quality of the prediction of the neural network.
We have found very good results using this simple
loss function and did not witness a significant smearing as reported 
elsewhere \citep[e.g.,][]{ledig16,schawinski17}.
The loss function is optimized with the Adam stochastic gradient descent type algorithm \citep{adam14} with a 
learning rate of $3 \times 10^{-4}$. The network is trained for 180 epochs with a batch size 
of 120\footnote{Training with stochastic gradient descent algorithms consists of iterating over the 
training set in batches. After each batch has been considered, the model parameters are updated
with the "noisy" estimation of the gradient. An epoch is an iteration over the whole
training set.}. The number of trainable parameters is 3.15 million.
Each epoch lasts for roughly 6 min on
an NVIDIA Titan X GPU, so the total training time is close to 19 hours. We 
check during training that the network is not
overtraining (adapting to the training dataset and not correctly generalizing)
by computing the loss on a validation set of images.

The recurrent neural network is definitively more difficult to train because
of the presence of recurrent connections, that slow down the backpropagation. Anyway, 
the residual connections and the Adam optimizer 
are enough to reach a good convergence. Following \cite{WieHirSchLen17}, the training is 
carried out by optimizing the following loss function:
\begin{equation}
  L = \sum_{i=1}^\mathcal{N} \sum_{j=2}^n \norm{\hat{I}_{(j)}(i) - I_{D}(i) }^2,
  \label{eq:loss_recurrent}
\end{equation}
where $n$ is the number of frames included during training (7 in our case, but
can be applied for any number of frames). This loss function forces that 
intermediate deconvolved frames approximate to
the final target. Summarizing, this scalar loss
function measures the mismatch between the target image and the prediction
produced by the recurrent neural network after each new frame is injected.
The number of trainable parameters in this case is 4.02 million.
Each epoch lasts for roughly 30 min on an 
NVIDIA P100 GPU, so the total training time is close to 2.5 days.

\subsection{Validation}
The results of the network applied to some patches of the validation set for the 
encoder-decoder arquitecture are displayed in Fig. \ref{fig:training_set}. The first
column shows the target image given by MOMFBD. The second column gives
the output of the deep neural network. For comparison, we display the best and
worst individual short-exposure frames (in terms of the rms contrast) in the last two columns 
and the time average of the
burst in the third column. It is obvious that the neural network approach is
able to extract high-frequency information from the burst of images, even though
this information is not present in each individual raw frame.

\begin{figure*}
  \centering
  \includegraphics[width=0.44\textwidth]{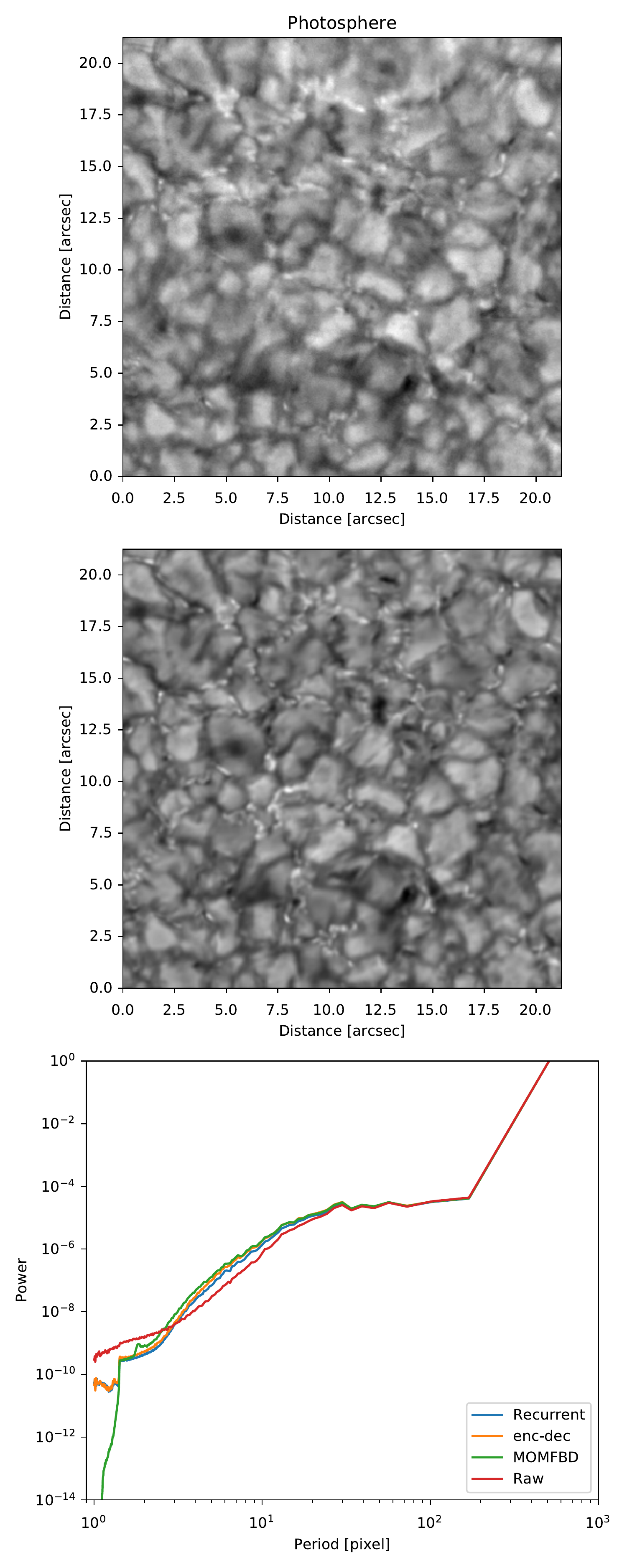}
  \includegraphics[width=0.44\textwidth]{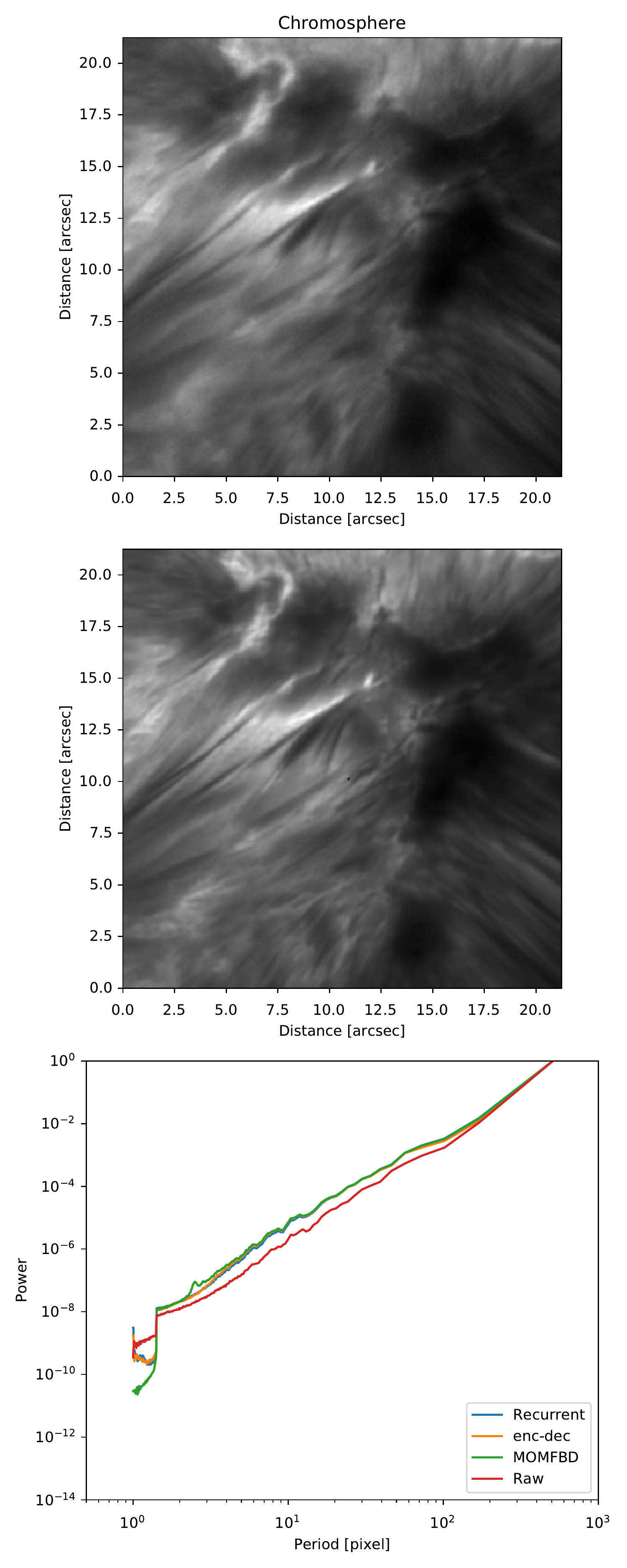}
  \caption{Top panels: a single raw image from the burst. Middle panels: reconstructed frames with the recurrent network. Lower
  panels: azimuthally averaged Fourier power spectra of the images. The left column shows results
  from the continuum image at 6302 \AA\, while the right column shows the results at the core of 
  the 8542 \AA\ line. All power spectra have been normalized to the value at the largest period (but
  not shown in the plot because it is outside the displayed range).}
  \label{fig:power_spectra}
\end{figure*}

\section{Results}
\subsection{Testing}
Once both neural networks are trained, we apply them to datasets
different to those used for training and validation. Given the fully convolutional
character of both architectures, one can apply it to frames of arbitrary
size (multiple of 8 to end up with input and outputs of the same size).
Additionally, this very same fully convolutional character allows the
network to transparently deal with spatially variant seeing conditions.
The correction carried out to each pixel in the input image is done as
if a potentially different PSF is perturbing it.
For this reason, there is no gain in correcting images by mosaicking and
one can input the full frames into the networks. The only practical limitation 
is in terms of memory, specially relevant for GPUs with their limited amount
of memory. One needs to make sure that the parameters of the networks, together
with the intermediate results of each step in the network can be allocated
in memory. We have not had any problem in deconvolving 1k$\times$1k images
in an NVIDIA Titan X GPU with 12 GB of memory. Anyway, images that do not
fit in memory can be corrected by mosaicking with some overlap and then stitching the patches
using, for instance, the median value for the overlapping regions. In terms of 
computing time, a Titan X GPU is able to deconvolve images of 1k$\times$1k in less than
5 ms for the encoder-decoder architecture and of the order of 50 ms
for the recurrent architecture.

\begin{figure}
  \centering
  \includegraphics[width=0.79\columnwidth]{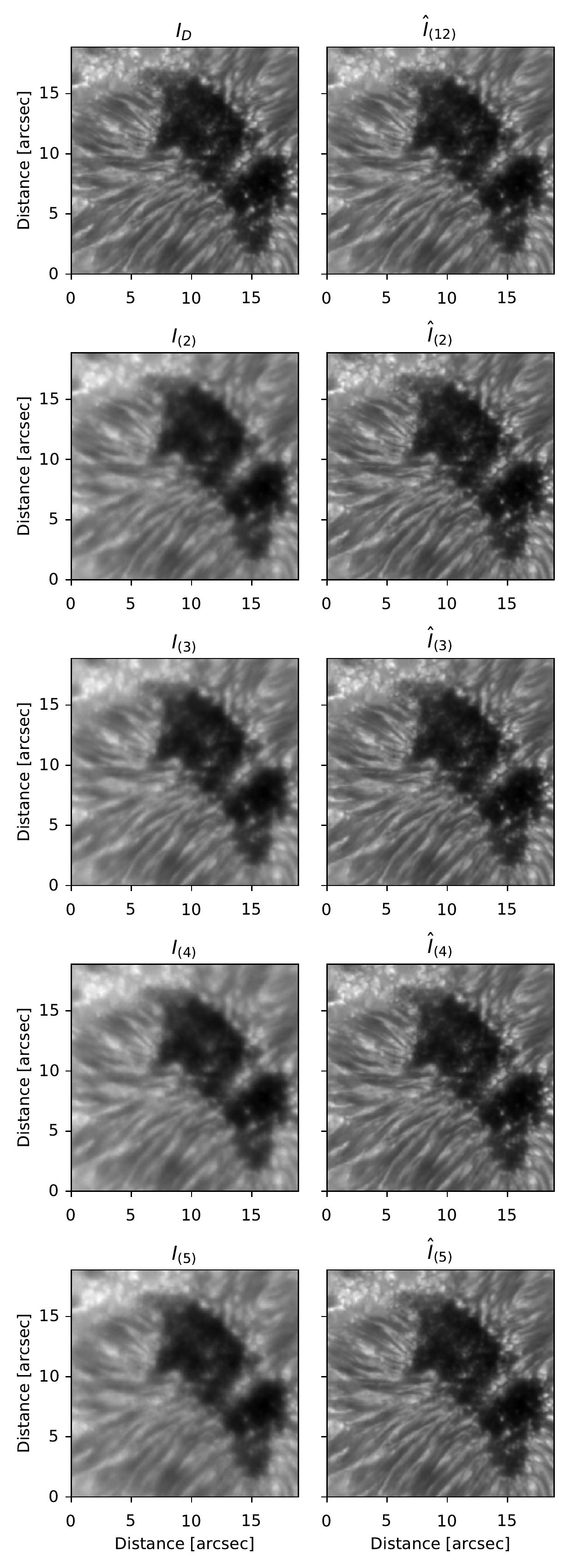}
  \caption{The top row displays the comparison of the 
  monochromatic image at the core of the Fe \textsc{i} 6173 \AA\ spectral line
  for MOMFBD (left), equivalently $I_D$ in out notation, and the output of the
  recurrent network after all the 12 frames are considered (right), equivalently, $\hat{I}_{(12)}$. The 
  next rows show the comparison between individual frames before and after the network 
  reconstruction. Note that we start from frame 2, where we have
  already used the first two observed frames.}
%  \caption{Reconstruction with the recurrent network for two different
%  wavelengths (left two columns for .... and right two columnd for ...
%  \textcolor{red}{ADUR cuales??}).}
  \label{fig:otherobs}
\end{figure}

Another interesting side effect of our learning-based approach
to solar image deconvolution is that no WB channel is, a priori, necessary anymore
for the deconvolution. The 
WB channel, taken in strict simultaneity with the monochromatic images
of the spectral scan and the polarimetric modulation, are used by the
MOMFBD algorithm to help in the deconvolution. Once the networks are
trained, this channel is not needed anymore, which largely facilitates
the instrumental setup. As a caveat, this is probably an exaggeration because
the WB channel is used in the current CRISP instrumental setup for other
purposes like image alignment.

% \textcolor{red}{We have used H$\alpha$ observations from the same target, which 
% is an interesting test because the wavelength of the observations is different than 
% those used in the training sets.} 

In terms of performance, Figs. \ref{fig:6302_validation} and \ref{fig:8542_validation} 
show the comparison between 
the encoder-decoder and the recurrent networks and the target MOMFBD image for two different 
monochromatic images, representative of the general behavior. The left column
shows the image obtained after the MOMFBD processing, with the upper row displaying an
image in the continuum and the lower row an image in the core of the lines.
In general, both networks produce a very similar output, with very detailed
fine structure. One can argue that there is an apparent lack of 
sharpness in the output of the networks as compared with the MOMFBD image. It is 
more obvious in the filamentary structure in the upper panel of Fig. \ref{fig:6302_validation}
and specially in the lower panel of Fig. \ref{fig:8542_validation}. Fibrils in
the core of the Ca \textsc{ii} 8542 \AA\ line are slightly more well defined in the
MOMFBD image than in the outputs of the networks. However, we think that part of the
sharpness is a consequence of the residual noise in the MOMFBD image, that appears as 
a strongly spatially correlated
structure, more visible in the upper panel of Fig. \ref{fig:8542_validation}.
It is clear that both networks are perfectly able to filter out this noise (that
is present in the training set). The random selection of patches for 
building the training set has the desirable consequence of breaking a significant part of the
spatial correlation of the noise in the MOMFBD images. Consequently, the networks are unable 
to reproduce it and, as a result, partially filter it out from the predictions.

These properties are better quantified by analyzing the spatial frequencies 
in the neural network predictions as compared with 
those of the original frames and the MOMFBD images. The left column of 
Fig. \ref{fig:power_spectra} shows the case of the continuum at 6302 \AA\ in a quiet Sun region
while we display the results for the core
of the 8542 \AA\ line in the right column of the same figure. The upper panel shows one
of the original frames, the middle panel displays the frame corrected with the
recurrent network while the lower panel shows the azimuthally averaged power spectrum
for the original and reconstructed frame with all the methods considered in this
work. The general behavior of the 
power spectrum is fundamentally the same for a large variety of observed frames and regions, so the
discussion can be made with only these two. Concerning the photospheric data, the 
individual frames present a clear
lack of power on spatial periods of $\sim$3 and $\sim$30 pixels, which correspond
to those mainly affected by the seeing. Both the recurrent end encoder-decoder arquitectures
are able to recover these spatial frequencies and increase their power, imitating
what is done with MOMFBD. Additionally, noise appearing in small scales is efficiently dampened 
by all image reconstruction methods. Concerning the chromospheric data,
we witness a small general increase in the power for almost all frequencies, except
in the very small scales, with periods below 2 pixels, where noise starts
to dominate over the signal.

\begin{figure*}
  \centering
  \includegraphics[width=0.70\textwidth]{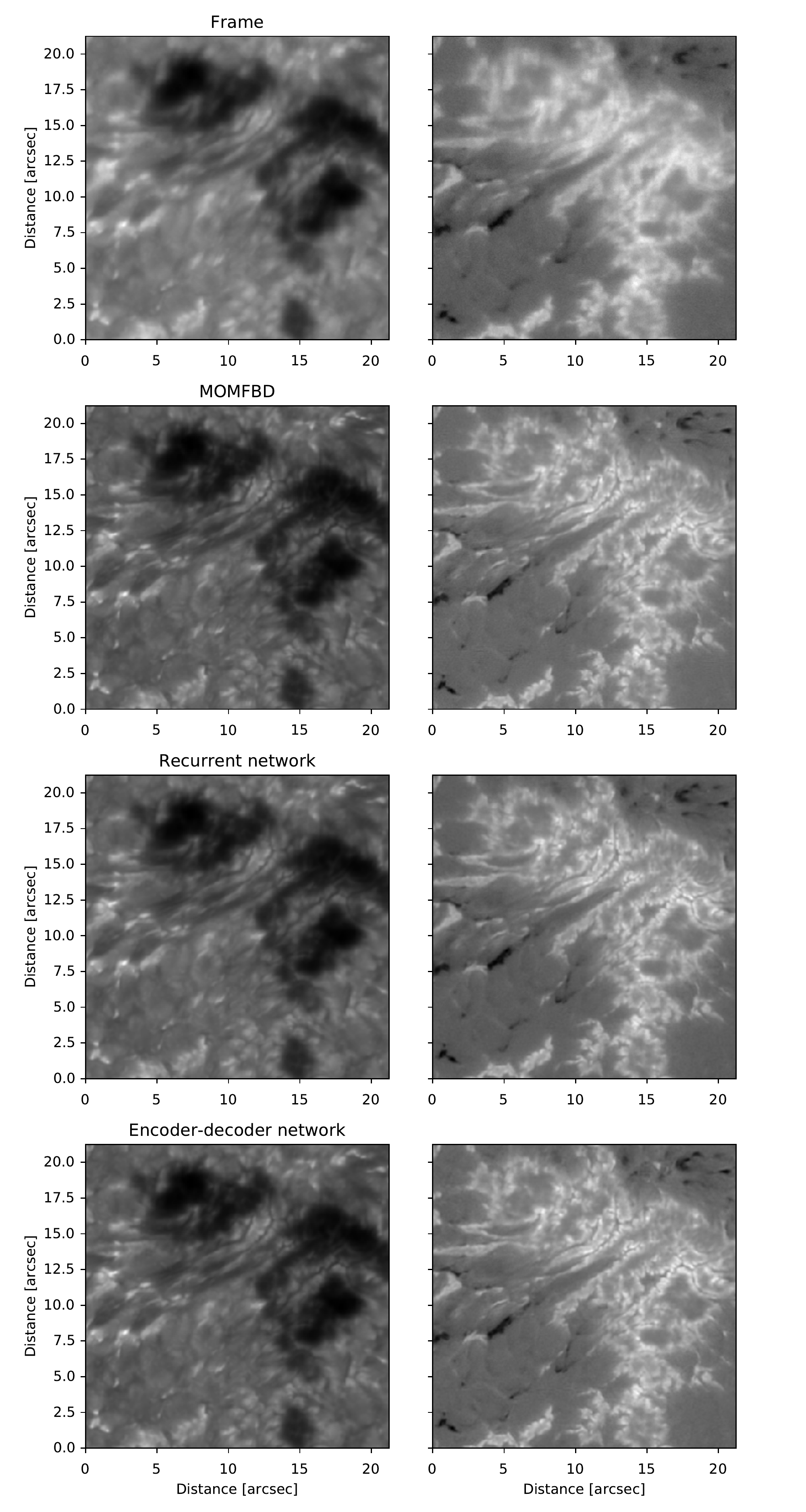}
  \caption{Left panels: monochromatic images in the core of the Fe \textsc{i} 6302 \AA\ line.
  Right panels: demodulated circular polarization signals for a single short-exposure
  frame, MOMFBD and the two architectures we propose in this work.}
  \label{fig:demodulation_6302}
\end{figure*}

\begin{figure*}
  \centering
  \includegraphics[width=0.70\textwidth]{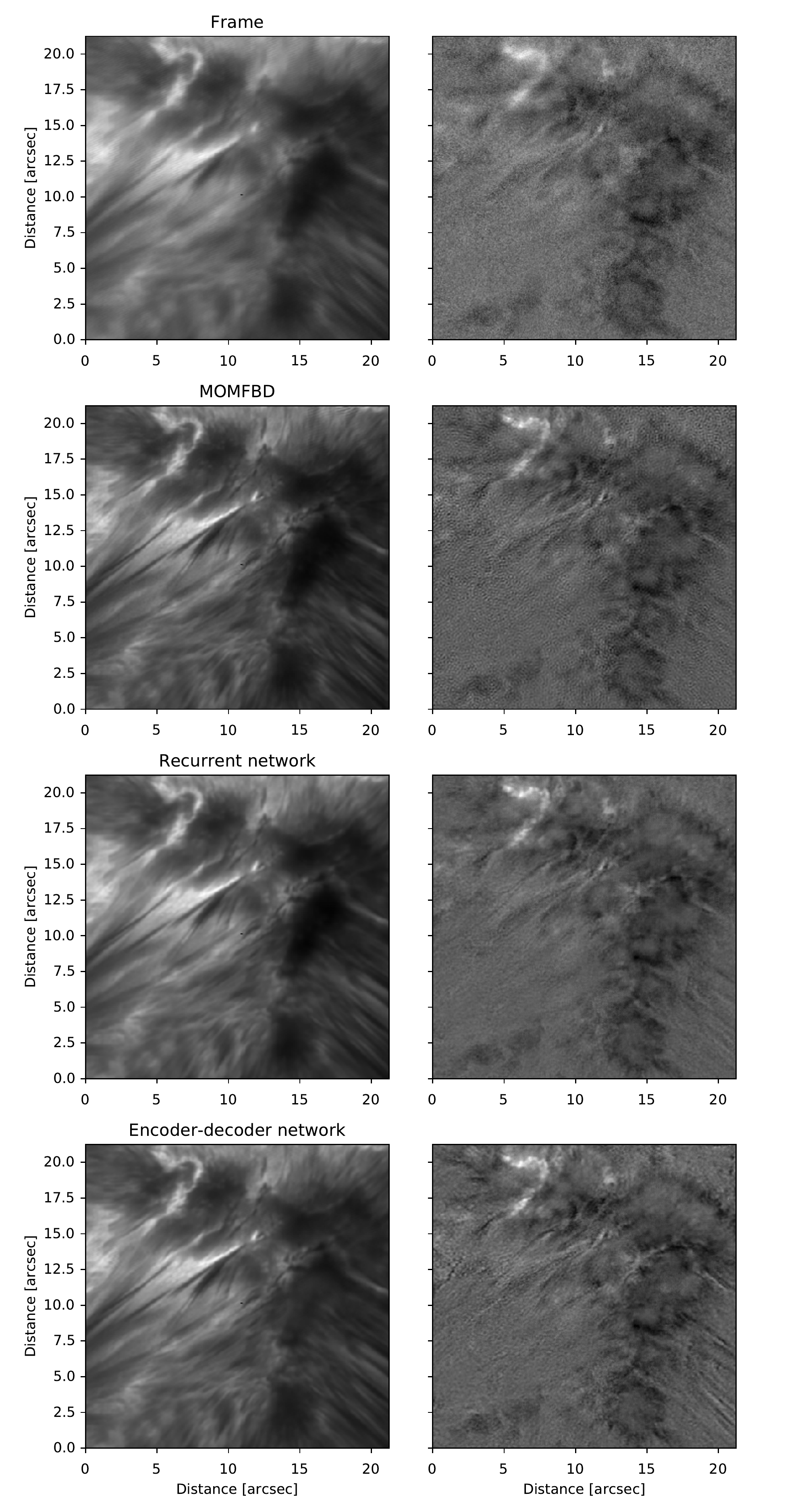}
  \caption{Same as Fig. \ref{fig:demodulation_6302} but for the Ca \textsc{ii} 8542 \AA\ line.}
  \label{fig:demodulation_8542}
\end{figure*}

\begin{figure*} 
  \centering
  \includegraphics[width=0.80\textwidth]{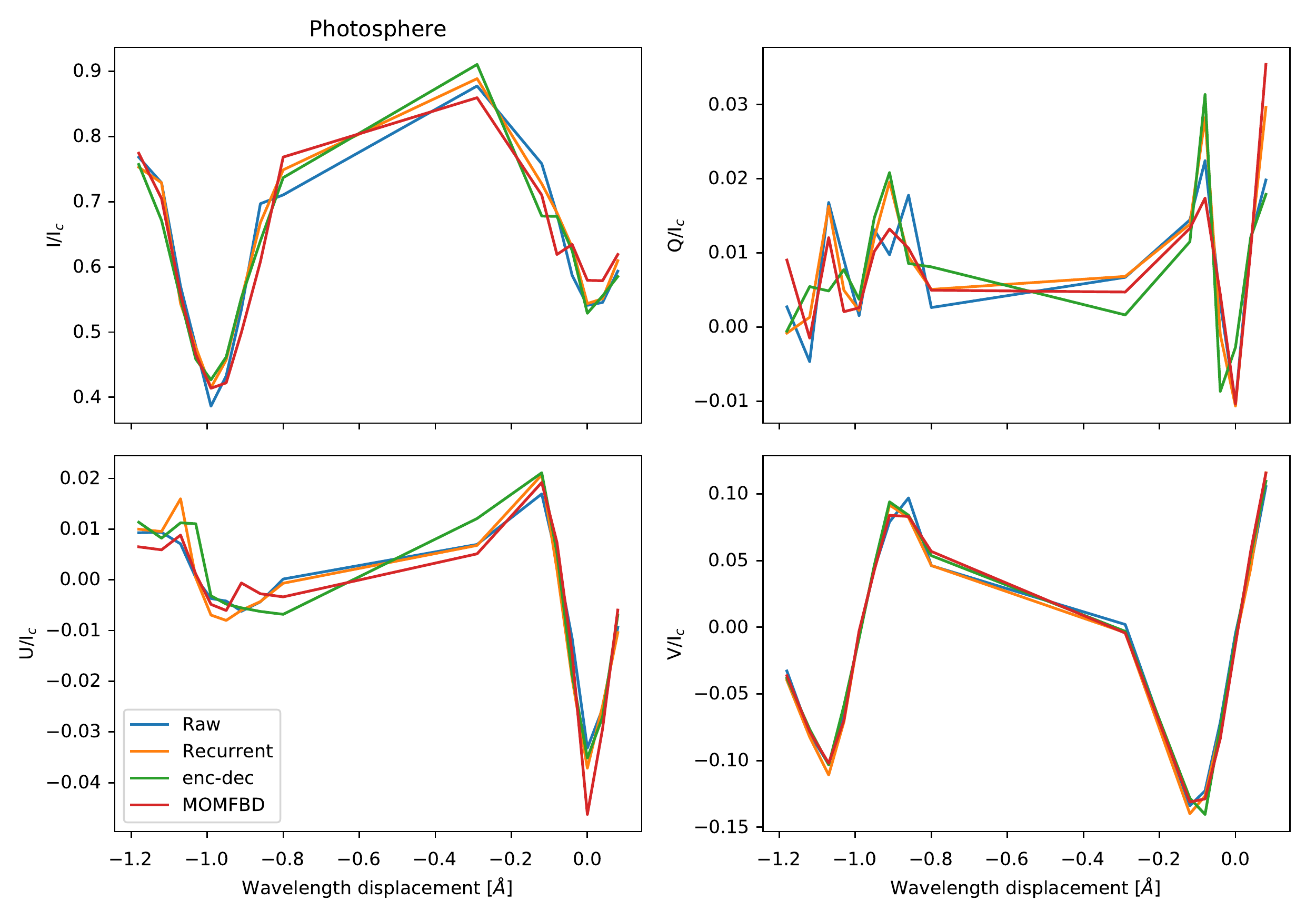}
  \includegraphics[width=0.80\textwidth]{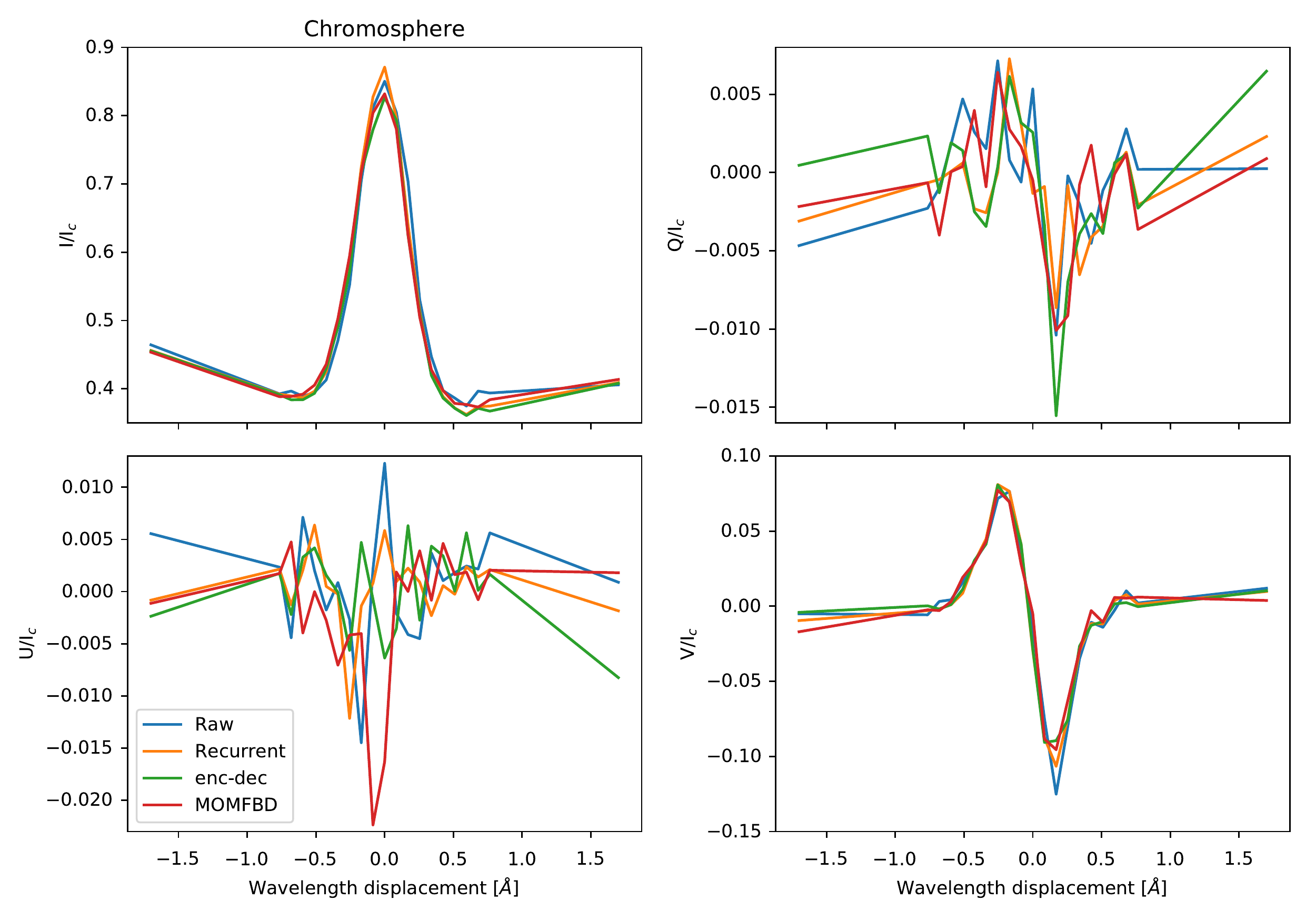}
  \caption{Four Stokes profiles for the same individual pixel both in the photosphere
  and chromosphere. We compare the original raw frame, together with the profile
  obtained after deconvolving with MOMFBD and the two architectures proposed in this work.}
  \label{fig:profiles}
\end{figure*}

To show the ability of the recurrent network to correct individual
frames, we show in Fig. \ref{fig:otherobs} the case of observations of 
the leading sunspot of AR12326 observed on 2015-04-19 using 
CRISP mounted on the SST. Two spectral regions were 
recorded sequentially: 6173 \AA\ (with the well-known Fe \textsc{i} line)
and 8542 \AA, acquiring four modulation     
states in 20 and 21 wavelength points, respectively. For each modulation 
state and spectral point, 12 (6) images were recorded in the 6173 \AA\ 
(8542 \AA) spectral range. In the top row of Fig. \ref{fig:otherobs}, we display the final 
MOMFBD reconstructed image, together with the final output of the recurrent 
network after all 12 frames have been considered. Additionally, we also 
show the individual frames (on the left) together with their reconstruction (right). Note
that all reconstructions have cumulative information from all previous frames. Note that the first 
reconstructed frame happens after two frames have been considered. 
It is clear that all individual reconstructed
frames are improved versions of the raw frames, even in the first frame.
This is a direct consequence of Eq. (\ref{eq:loss_recurrent}), which forces
all individual frames to converge towards the final corrected frame. Note in passing
that the Fe \textsc{i} 6173 \AA\ line was not included in the training. These results
show a first glimpse at the generalization ability of the trained networks, which
is expanded in Sect. \ref{sec:halpha}.

\subsection{Polarimetric demodulation}
As mentioned in Section \ref{sec:training_set}, the training and validation
sets contain modulated Stokes parameters. Given that the linear and
circular polarization signals in the Sun are often very small (normally 
much smaller than a percent in units of the continuum intensity except
in strongly magnetized regions), any deconvolution process might run the risk
of creating artificial signals. The reason is that polarimetric signals
are obtained after combining several (typically four) observed frames. If
during the deconvolution we introduce differential corrections in 
these frames, spurious signals will be created on the demodulation. To check for this issue, we
display in Fig. \ref{fig:demodulation_6302} the demodulated circular polarization 
signals in the wing of the Fe \textsc{i} 6302 \AA\ line. Likewise, Fig. \ref{fig:demodulation_8542}
shows the demodulated Stokes $V$ signals in the wing of the Ca \textsc{ii} 8542 \AA\ line.
A direct comparison of the four magnetograms in Fig. \ref{fig:demodulation_6302} 
shows that the same fine structure
is present in all of them, without any visible spurious signal appearing in any
of the two neural architectures. It is obvious that the fine structure of the 
monochromatic image and the magnetogram are nicely recovered by our neural
network approach, concentrating the polarization signals that were 
smoothed out by the atmosphere.

\begin{figure*}
  \includegraphics[width=\textwidth]{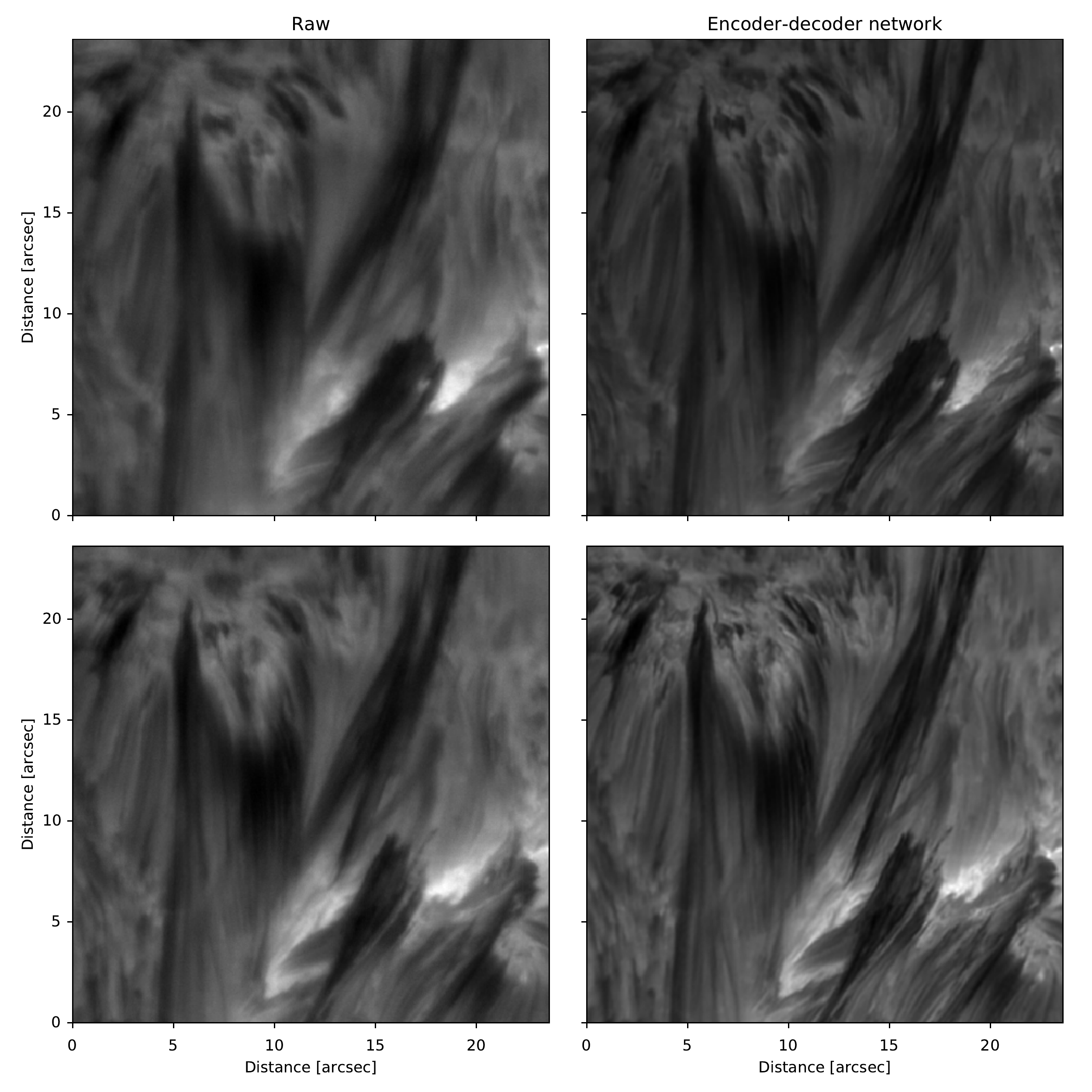}
  \caption{Individual raw frames (left column) and deconvolved images (right column) for
  two different time steps of monochromatic images in the core of the H$\alpha$ line.}
  \label{fig:halpha}
\end{figure*}

There are some differences, though, in the 
weaker signals of the Ca \textsc{ii} 8542 \AA\ line. The recurrent network
seems to produce slightly better results than the encoder-decoder architecture. The
clearly visible spatially correlated noise in the MOMFBD reconstruction is strongly
dampened by the neural network approach, producing much cleaner magnetograms.
Also, the spurious (at least not present in MOMFBD) structures that appear
more conspicuously
in the region around (20'',2.5''), which are specially visible in the encoder-decoder
network, are not recovered by the recurrent architecture. These elongated structures seem to be 
a consequence of an artificial cross-talk between Stokes $I$ and $V$ produced by
differential corrections in the different frames used by the polarimetric demodulation.
A possible solution to these artifacts is proposed in Sect. \ref{sec:shuffle}. An issue, that
according to our experiments apparently only happens in the encoder-decoder architecture, 
is that the contrast of the umbrae is reduced
with respect to the surroundings.

It is also interesting to check the effect of the deconvolution on
the Stokes profiles of an individual pixel. To this end, we display
in Fig. \ref{fig:profiles} the Stokes profiles of a single pixel in 
both spectral lines. As reference, we show the Stokes profile of one
of the frames of the burst in blue. We point out that the monochromatic
deconvolution carried out by MOMFBD and the neural architectures proposed
in this work do not produce any sizable artifact in these pixels with
strong polarimetric signals. Pixels with weaker signal are, however, more affected
by noise.

\subsection{H$\alpha$ monochromatic images}
\label{sec:halpha}
A neural network that is correctly trained should be generalizable
to other inputs, providing reliable outputs. To this end, we tested
the two networks developed in this work to images in the core of 
the H$\alpha$ line. The core of the line displays very dynamic fibrilar structures
that are believed to be tracing chromospheric material \citep{rutten08},
although its physics is still not fully understood \citep{leenaarts15,rutten17}.
Some of the fibrils present in monochromatic H$\alpha$ images are also partially seen 
in the core of the Ca \textsc{ii} 8542 \AA\ line, so one should expect
the neural networks to generalize well. We use data obtained
on 2016-09-19 from 09:30 to 10:00 with CRISP on the SST. The rows of Fig. \ref{fig:halpha}
shows two consecutive frames in the core of H$\alpha$, with one of the original
frames in the first column and the deconvolved image using the
encoder-decoder arquitecture in the second column. We do not display
the results of the recurrent architecture to maximize the size of the
images for a better comparison. The results are indeed very similar
with the two neural networks. It is clear from these results that the network has been
able to learn the process of image deconvolution, revealed by the huge amount 
of small-scale substructure appears in the H$\alpha$ 
filaments\footnote{A movie showing the time evolution of H$\alpha$ monochromatic
images can be found on the repository for the code.}.

\begin{figure*}
  \centering
  \includegraphics[width=0.65\textwidth]{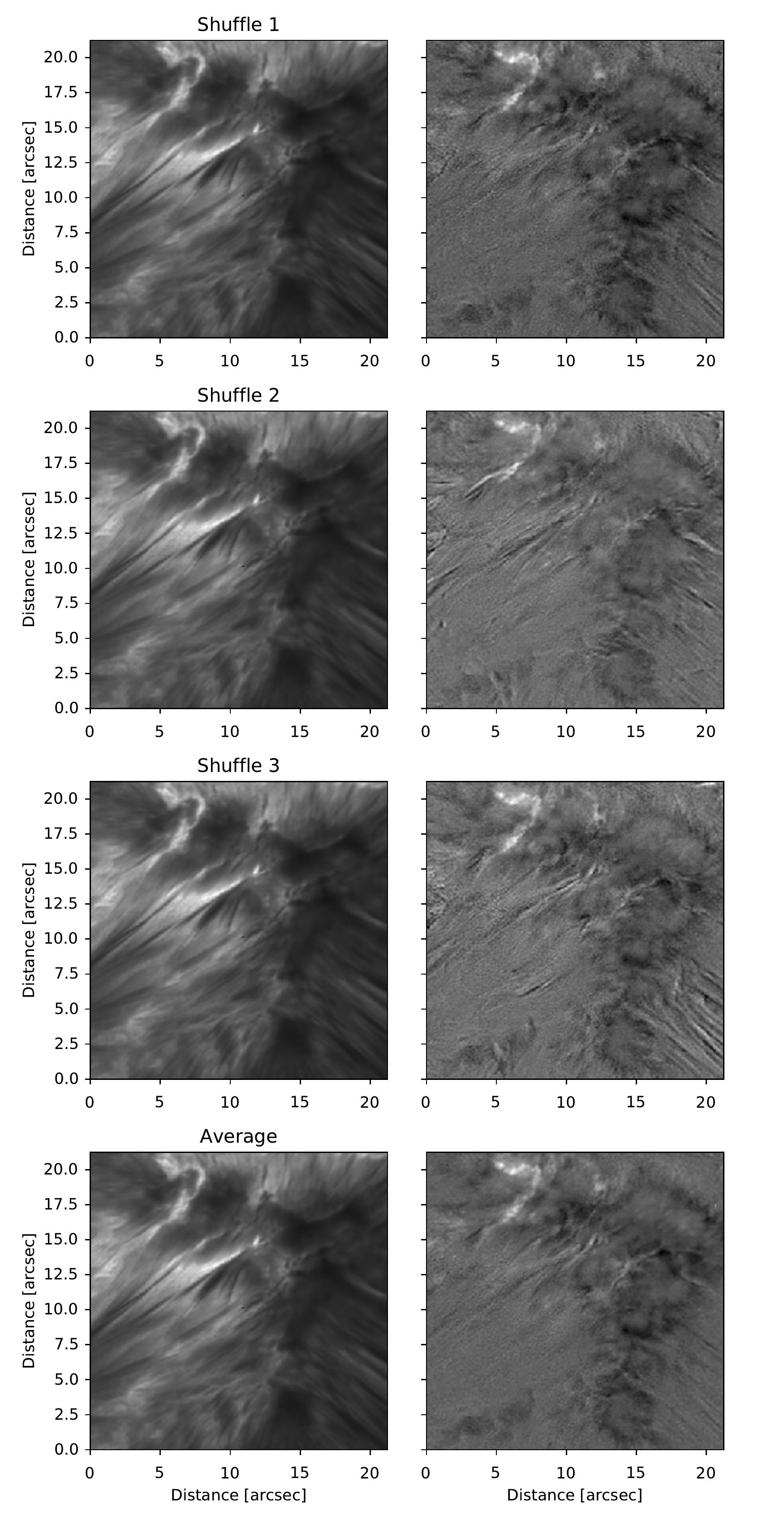}
  \caption{Deconvolved images in the core of the Ca \textsc{ii} 8542 \AA\
  line (left column) and demodulated circular polarization (right column)
  for different shuffling of the input using the encoder-decoder network.
  The final row shows the average of 20 different shufflings.}
  \label{fig:shuffle}
\end{figure*}

\subsection{Frame shuffling and committees}
\label{sec:shuffle}
All the experiments carried our so far use the input frames in the order
obtained in the telescope. Changing the order of the input by reshuffling
can be understood as another check for overtraining to verify that the networks 
have not memorized the ordering imposed during training. Additionally, the 
different outputs can be later combined following the neural network committee machine paradigm
to improve the output. Given the nonlinear denoising character of the network, 
it makes sense to combine the
outputs to improve the signal-to-noise ratio and reduce artifacts. This is
indeed what happens, according to Fig. \ref{fig:shuffle}, with the 
encoder-decoder network (the same happens in the recurrent architecture). The fundamental
reason for this behavior in the encoder-decoder architecture is that the first convolutional block, $C_{0,1}$, 
produces images with 64 channels by combining with different weights the 7 input frames. This information
is then propagated to the output and, as a consequence, a shuffling in the input layer produces 
a different output. A similar explanation is the source of the variability in the recurrent network.
The first three
rows of the figure displays three different outputs by shuffling the input images in the core of 
the Ca \textsc{ii} 8542 \AA\ line in the left column and in the peak of
Stokes $V$ in the right column. It is clear that different shufflings produce
different artifacts, specially in the lower right part of the image. This demonstrates
that they are indeed produced by non-perfect reconstructions, which produce
artificial polarization signals after demodulation. The last row
gives the simplest committee machine one can think of, by plain averaging
20 such shufflings. More elaborate committees can be studied in the future. 
The spatial resolution of the final averaged image is \emph{practically} the same as that of 
Fig. \ref{fig:demodulation_8542} because the real signals all appear at practically the same position and 
scale in all instances of the commitee. However, the artifacts are different in all instances 
and tend to average to zero. Of special relevance is the disappearance 
of the fibrilar structures in the lower right corner on the circular polarization maps.
Additionally, we witness a small decrease on the noise variance (around a factor 2), 
probably a mixture of artifact reduction plus a more efficient denoising.

\section{Conclusions}
We propose in this paper an end-to-end approach for multiframe
blind deconvolution of solar images based on deep convolutional
neural networks. We have used two different architectures. The first one
is a relatively simple encoder-decoder network which needs to fix a-priori
the number of frames considered. The second one is a recurrent 
architecture that can work with an arbitrary number of frames. Both
architectures are trained with data from CRISP@SST and provide very 
fast image reconstructions.

We have demonstrated that the neural networks generalize well to 
unseen data. They are also able to keep the photometric quality of the
data without compromising the modulated signals, thus providing polarimetry data 
of high quality. Additionally, the
networks produce comparable results to the MOMFBD reconstruction without
making use of simultaneous WB images. The only information 
needed for our end-to-end solution are the monochromatic frames
produced by the etalon narrow filter.

Both networks are made available to the 
community, with all the training and testing details necessary for the reproduction of the results
presented in this paper. The networks are developed with \texttt{PyTorch}
which make very efficient use of any available GPU. Although the image correction is still not
done on real time with current hardware for large images of 1k $\times$ 1k, 
we expect that future improvements on hardware and neural networks 
architectures that one can achieve after some ablation studies allow
this approach to deal with images in real time. Anyway, the networks
can be easily deployed on the telescope and produce quasi-realtime
image reconstruction. Even if our approach is not used as the final
processing for the science-ready data or managed to run in real-time, they will produce 
very valuable information that can be 
used by the observer to have a peek on how will science data
look like at the end of the MOMFBD processing. 

% Another interesting
% avenue to investigate is the development of neural networks for the estimation
% of the per-pixel wavefront.

Despite the successful neural networks presented here, we 
anticipate that improvements on the quality of the reconstruction can
be achieved. First, one can use more elaborate loss functions. It is known that
the $\ell_2$ norm of the residual tends to produce fuzzy reconstructions,
specially when the number of frames is small. One can achieve better
results by utilizing adversarial training \citep[e.g.,][]{ledig16}. Second, every burst of
images is processed in isolation, so there is no transfer of information from burst to burst.
Imposing some kind of time consistency can help the neural networks successfully correct
burst in which the seeing get worse. One can anticipate that a hierarchical
architecture in which information is propagated for consecutive frames and also for consecutive
bursts (similar to our recurrent architecture but with another level of hierarchy) 
might be worth trying. Finally, 
we note that training DNNs for multiframe image deconvolution can potentially be done
using synthetic data. To this end, one could generate synthetic images from available 
magneto hydrodynamic simulations of the solar atmosphere and perturb them using
artificial wavefronts from a synthetic turbulent atmosphere. The advantage
in the synthetic case is that the unperturbed image is available to us, and
the application of the MOMFBD algorithm is not needed. This can help
reducing some of the artifacts produced by the almost unavoidable presence of spatially 
correlated noise in the MOMFBD data.
However, this approach might suffer from
some lack of realism and it is not yet clear whether this approach will generalize correctly. 
It is currently very difficult to carry out simulations 
with a sufficiently realistic chromosphere. Therefore, the synthetic spectral lines
are still not as realistic as desired. Moreover, computing the effect of a sufficiently
realistic turbulent atmosphere is a hard task. One needs to take into account
the known temporal correlation in the wavefront deformation and the presence of
multiple anisoplanatic patches covered by the images.
Work along these line is presumably needed in the
future.

% \begin{figure*}
%   \includegraphics[width=0.45\textwidth]{ca8542_power_wave_10.pdf}
%   \includegraphics[width=0.45\textwidth]{ca6302_power_wave_10.pdf}
% \end{figure*}

\begin{acknowledgements}
We thank Michiel van Noort for useful suggestions and the referee for many
interesting suggestions that improved the paper. Financial support by the Spanish Ministry of Economy and Competitiveness 
through projects AYA2014-60476-P. We also thank the NVIDIA Corporation for the donation
of the Titan X GPU used in this research.
This research has made use of NASA's Astrophysics Data System Bibliographic Services.
JdlCR is supported by grants from the Swedish Research Council (2015-03994), the Swedish 
National Space Board (128/15) and the Swedish Civil Contingencies Agency (MSB). 
This project has received funding from 
the European Research Council (ERC) under the European Union's Horizon 2020 research and 
innovation programme (SUNMAG, grant agreement 759548). 
APY is supported by the German Government through DFG project ``STOK3D Three dimensional Stokes 
Inversion with magneto-hydrostationary constraints''. 
The Swedish 1-m Solar Telescope is operated on the island of La Palma by the Institute for 
Solar Physics of Stockholm University in the Spanish Observatorio del Roque de los Muchachos 
of the Instituto de Astrof\'isica de Canarias. The Institute for Solar Physics is supported
by a grant for research infrastructures of national importance from the Swedish Research Council (registration
number 2017-00625). 
This study has been discussed
in the workshop \emph{Studying magnetic-field-regulated heating in the solar chromosphere} (team 399) 
at the International Space Science Institute (ISSI) in Switzerland.
We acknowledge the community effort devoted to the development of the following open-source packages that were
used in this work: \texttt{numpy} (\texttt{numpy.org}), 
\texttt{matplotlib} (\texttt{matplotlib.org}), 
and \texttt{PyTorch} (\texttt{pytorch.org}).
\end{acknowledgements}

% \bibliographystyle{aa}
% \bibliography{biblio}
% apy start
%\bibliography{/scratch/Dropbox/biblio}
% apy end

\end{document}